\documentclass[10.5pt,letterpaper,fleqn, twocolumn]{article}

\usepackage{graphicx}
\usepackage{makeidx}
\usepackage{multicol}
\usepackage{crop}
\usepackage{amstext,amsthm}
\usepackage{amsmath,amssymb}
\usepackage{bm}
\usepackage{natbib}

\newtheorem{theorem}{Theorem}[section]

\newtheorem{remark}[theorem]{Remark}

\usepackage{multicol}
\usepackage{pslatex}
\usepackage[strict]{changepage}
\newcommand{\be}{\begin{equation}}
\newcommand{\ee}{\end{equation}}
\newcommand{\bea}{\begin{eqnarray}}
\newcommand{\eea}{\end{eqnarray}}
\newcommand{\bne}{\begin{equation*}}
\newcommand{\ene}{\end{equation*}}
\newcommand{\bi}{\begin{itemize}}
\newcommand{\ei}{\end{itemize}}

\newcommand{\bbm}{\begin{bmatrix}}
\newcommand{\ebm}{\end{bmatrix}}

\newcommand{\mr}{\mathrm}

\usepackage[top=0.75in, bottom=0.5in, left=0.83in, right=0.83in]{geometry}
\usepackage[hang,splitrule]{footmisc}





\usepackage{titlesec}		
\usepackage{subcaption}		
\usepackage{indentfirst}	

\newenvironment{myabstract}
{
	\vspace{0.2in}
	\parindent=0cm \textit{Abstract}
	\parindent=0.5cm \hangindent=0.5cm \linebreak
}

\newenvironment{mykeywords}
{
	\vspace{0.2in}
	\parindent=0cm \textit{Keywords}
	
	\parindent=0.5cm
}

\titleformat{\section}
{\normalfont\bfseries}
{}{0.0pt}{\parindent=0cm}[\parindent=0.5cm]

\titlespacing*{\section}
{0pt}{1.5\baselineskip}{0.5\baselineskip}

\titleformat{\subsection}
{\normalfont\itshape}
{}{0.0pt}{\parindent=0cm}[\parindent=0.5cm]

\titlespacing*{\subsection}
{0pt}{1.0\baselineskip}{0.5\baselineskip}

\begin{document}

\twocolumn[{%
	
\phantom\\
\vspace{0.5in}
\begin{center}
\Large{\textbf{Model-based control algorithms
for the quadruple tank system:\\ An experimental comparison}}\\
\end{center}
\vspace{0.2in}

\begin{center}
Anders H. D. Andersen $^{\mr{a}}$, Tobias K. S. Ritschel $^{\mr{a}}$, Steen Hørsholt $^{\mr{a}}$\\ Jakob Kjøbsted Huusom $^{\mr{b}}$ and John Bagterp Jørgensen $^{\mr{a,}}$\footnote \\

\vspace{0.10in}

$^{\mr{a}}$ Department of Applied Mathematics and Computer Science,\\
Technical University of Denmark, DK-2800 Kgs. Lyngby, Denmark

\vspace{0.10in}
$^{\mr{b}}$ Department of Chemical and Biochemical Engineering,\\
Technical University of Denmark, DK-2800 Kgs. Lyngby, Denmark

\end{center}

\begin{myabstract}
	We compare the performance of proportional-integral-derivative (PID) control, linear model predictive control (LMPC), and nonlinear model predictive control (NMPC) for a physical setup of the quadruple tank system (QTS). 
We estimate the parameters in a continuous-discrete time stochastic nonlinear  model for the QTS using a prediction-error-method based on the measured process data and a maximum likelihood (ML) criterion. In the NMPC algorithm, we use this identified continuous-discrete time stochastic nonlinear model. The LMPC algorithm is based on a linearization of this nonlinear model. We tune the PID controller using Skogestad's IMC tuning rules using a transfer function representation of the linearized model. Norms of the the observed tracking errors and the rate of change of the manipulated variables are used to compare the performance of the control algorithms. The LMPC and NMPC perform better than the PID controller for a predefined time-varying setpoint trajectory. The LMPC and NMPC algorithms have similar performance.

\end{myabstract}

\begin{mykeywords}
    Quadruple Tank System, PID Control, Linear MPC, Nonlinear MPC, SysID, Experimental Comparison
\end{mykeywords}

\vspace{0.2in}
}]

\footnotetext[1]{\parindent=0cm \small{Corresponding author: J. B. J{\o}rgensen (E-mail: {\tt\small jbjo@dtu.dk}).}}

%
%



\section{Introduction}

In the process industries, advanced process control (APC) strategies are used to maximize profit by increasing operation efficiency and reducing process variability. Model predictive control (MPC) is a widely used APC methodology and numerous successful implementations have been reported in real industrial systems \citep{Bauer2008}. 
However, most process control loops still consist of proportional-integral-derivative (PID)-type control systems despite the inherently complex nature of industrial process systems \citep{aastroem1995}.

Compared to standard PID-type control strategies, the anticipatory behavior of the MPC methodology offers superior tracking capabilities of predefined time-varying setpoints for strongly interconnected multi-input multi-output systems.  Compared with linear MPC, nonlinear MPC can further improve setpoint tracking for systems where the nonlinear dynamics are significant~\citep{LMPCvsNMPC}. An MPC strategy requires a mathematical model of the process and any plant-model mismatch impacts the closed-loop performance. The performance of different control algorithms (e.g., PID and MPC algorithms) is usually compared using a test system. The quadruple tank system (QTS) is a classical example of such a test system, and several research papers describe simulation and experimental tests of PID and MPC strategies applied to the QTS \citep{KarlHenrikJohansson2000, QTS_MPC_sim_exp, Sazuan_MPC}. \cite{QTS_MPC_sim_exp} compare the performance of a PI-controller based system with an LMPC applied to a physical setup of the QTS. Compared to previous studies, the novelties in our paper are systematic system identification, the use of a model-based tuning procedure for the PID-controller based system, and systematically testing with a predefined time-varying setpoints trajectory allowing for anticipatory actions in the MPCs.\\ 
\indent
We present a comparative study of standard implementations of a PID controller, an LMPC, and an NMPC applied to a physical setup of the QTS. The NMPC involves the solution of an optimal control problem (OCP) with input constraints and a continuous-discrete extended Kalman filter (CD-EKF) for estimating states and unmeasured disturbances. Similarly, the LMPC is based on 1)~the solution of an OCP and 2)~a continuous-discrete Kalman filter (CD-KF) for estimating the states and unmeasured disturbances. We present a continuous-discrete time stochastic nonlinear model and we use it as the process model in the NMPC design. 
The parameters in the model used by the controllers are identified using a maximum likelihood (ML) prediction-error-method (PEM). The estimated model is used instead of a model with nominal parameters as this reduces the plant-model mismatch significantly. For the LMPC, we use a linearized version of this model as the process model.
We systematically tune the PID control system using Skogestad's IMC model-based tuning rules applied to transfer functions derived from the linearized version of the continuous-discrete time stochastic nonlinear model \citep{inbook_Skogestad}. 
Finally, we perform experiments using predefined time-varying setpoints for the two bottom tanks of the QTS for all three control algorithms. We use the data from these experiments to compare the performance of the PID, LMPC, and NMPC algorithms in terms of tracking errors and the rate of change in the manipulated variables (MVs).\\
\indent
The remaining part of this paper is organized as follows.
Section \ref{sec:modeling} presents the models. Section \ref{sec:estimation} describes the CD-EKF and the CD-KF, while Section \ref{sec:MLPEM} describes a prediction-error-method for parameter estimation. In Section \ref{sec:control}, we discuss the standard PID, LMPC, and NMPC algorithms used in this study, as well as the tuning of the controller parameters. Section \ref{sec:experiments} presents the data obtained from experiments performed on a physical setup of the QTS, and the control algorithms are compared using different norms of the observed tracking errors and rate of change of the manipulated variables. In Section \ref{sec:conclusion}, we present conclusions.

\section{Modeling}
\label{sec:modeling}
\begin{figure}[tb]
    \centering
    \centerline{\includegraphics[width=0.40\textwidth]{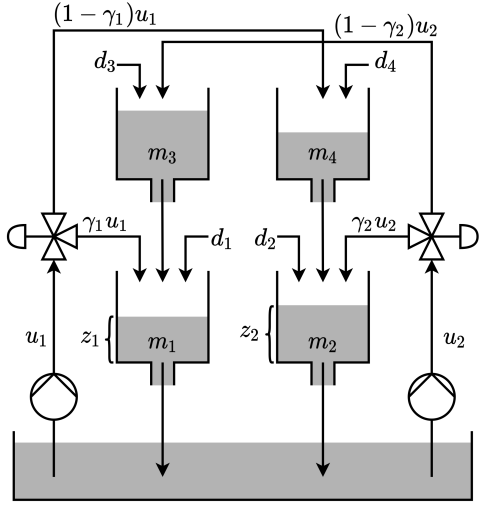}}
    \caption{Schematic diagram of the quadruple tank system.}
    \label{fig:QTS}
\end{figure}

The QTS consists of four water tanks, two valves, and two pumps, as shown in Figure \ref{fig:QTS}. Pump 1 fills tanks 1 and 4 and pump 2 fills tanks 2 and 3. Tank 4 discharges to tank 2 while tank 3 discharges to tank 1.  Valve 1 controls the fraction of water from pump 1 that flows into tank 1 and valve 2 controls the fraction of water flow from pump 2 into tank 2. 

We model the QTS as a nonlinear stochastic continuous-discrete system,
\begin{subequations}
\label{eq:model_structure}
\begin{alignat}{3}
        dx(t) &= f(x(t), u(t), d(t), \theta)dt+\sigma( \theta)d\omega(t),\label{eq:SDE}\\
        y(t_k) &= g(x(t_k),\theta)+v(t_k),\label{eq:meas_eq}\\
        z(t) &= h(x(t), \theta),\label{eq:CVs}
\end{alignat}
\end{subequations}
where $t$ is time, $x(t) = \begin{bmatrix}
m_1(t); & m_2(t); & m_3(t); & m_4(t)
\end{bmatrix}$ is the state vector representing the masses [g] of water in the tanks, $u(t) = \begin{bmatrix}
u_1(t); & u_2(t)
\end{bmatrix}$ are the MVs representing inflows [cm$^3$/s] from
the two pumps to the tanks, $y(t_k) = \begin{bmatrix}
y_1(t_k); & y_2(t_k); & y_3(t_k); & y_4(t_k)
\end{bmatrix}$ is a vector representing the measured water levels [cm] in the tanks, $z(t) = \begin{bmatrix}
z_1(t); & z_2(t)
\end{bmatrix}$ are the controlled variables (CVs) representing the water levels [cm] in the bottom tanks, $d(t) = \begin{bmatrix}
d_1(t); & d_2(t); & d_3(t); & d_4(t)
\end{bmatrix}$ are the disturbance variables representing plant-model mismatch in the form of unknown inflows [cm$^3$/s] in all the tanks, $\omega(t)$ is a standard Wiener process, i.e., $d\omega(t) \sim N_{iid}(0, Idt)$ [$\sqrt{\mathrm{s}}$], and $v(t_k) \sim N_{iid}(0, R)$ is a discrete-time independent and identically normally distributed stochastic process with covariance $R = \mathrm{diag}([r_{1}^2,\: r_{2}^2,\: r_{3}^2,\: r_{4}^2])$.
$\theta$ is the time-invariant parameter vector. The system of first-order stochastic differential equations in \eqref{eq:SDE} is the mass balances,
\begin{alignat}{1}
        d m_i(t) &= \left( \rho q_{i,in}(t) - \rho q_{i,out}(t) \right) dt + \sigma_{i} d\omega_i(t), 
        \label{eq:differential_quadruple_tank_system}
\end{alignat}
for $i \in \{1, 2, 3, 4\}$. 
$\rho = 1.0\:\mathrm{g}/\mathrm{cm}^3$ is the density of water.
The water flowing into the tanks are described by
\begin{subequations}
    \begin{alignat}{1}
    q_{1,in}(t) &= \gamma_1u_1(t)+d_1(t)+q_{3,out}(t),\\
            q_{2,in}(t) &= \gamma_2u_2(t)+d_2(t)+q_{4,out}(t),\\
            q_{3,in}(t) &= (1-\gamma_2)u_2(t)+d_3(t),\\
            q_{4,in}(t) &= (1-\gamma_1)u_1(t)+d_4(t),
    \end{alignat}
\end{subequations}
where $\gamma_1, \gamma_2 \in (0,1)$ represent the valve configurations. The water flowing out of the tanks are described as
\begin{alignat}{3}
        q_{i,out}(t) &= a_i\sqrt{2g_ah_i(t)}, \quad h_i(t) &&= \frac{m_i(t)}{\rho A_i},
\end{alignat}
for $i \in \{1, 2, 3, 4\}$ and where $g_a = 981\:\mathrm{cm}/\mathrm{s}^2$ is the acceleration of gravity \citep{KarlHenrikJohansson2000}. 
The CVs in \eqref{eq:CVs} are the water levels in the bottom tanks,
\begin{equation}
   h(x(t), \theta) = C_z(\theta)x(t), \quad C_z(\theta) =\begin{bmatrix}
 \frac{1}{\rho A_1}&0  & 0 & 0\\ 
0 &\frac{1}{\rho A_2}  &  0& 0
\end{bmatrix}.
\end{equation}
The measurement equation \eqref{eq:meas_eq} is the measured water levels in all tanks, i.e.,
\begin{equation}
    g(x(t_k), \theta) = C(\theta)x(t_k), 
\end{equation}
where
\begin{equation}
\label{eq:C_matrix}
    C(\theta) = \mathrm{diag}\bigg(\bigg[\frac{1}{\rho A_1},\: \frac{1}{\rho A_2},\: \frac{1}{\rho A_3},\: \frac{1}{\rho A_4}\bigg]\bigg).
\end{equation}
The diffusion coefficient in \eqref{eq:SDE} is modeled using a diagonal matrix,
\begin{equation}
    \sigma(\theta) = \mathrm{diag}([\sigma_{1},\: \sigma_{2},\: \sigma_{3},\: \sigma_{4}]).
\end{equation}
The nominal parameters of the QTS that represent the cross-sectional area of the outlet tubes and tanks, $a_i$ and $A_i$, are $a_i = 1.131\; \mathrm{cm}^2$ and $A_i = 380.133\; \mathrm{cm}^2$ for $i \in \{1, 2, 3, 4\}$. We choose the valve configurations as $\gamma_j = 0.35$ for $j \in \{1, 2\}$. Consequently, the QTS has non-minimum phase characteristics. We express these parameters together with the diffusion coefficients as the parameter vector $\theta$. 

We derive a linear model of the QTS by making Taylor approximations of the nonlinear stochastic continuous-discrete model at the operating point $(x_s, u_s, d_s)$,
\begin{subequations}
\label{eq:linear_state_space_model}
\begin{alignat}{3}
       dX(t) &= \bigg(A(\theta)X(t)+B(\theta)U(t)+E(\theta)D(t)\bigg)dt\nonumber \\
       &+\sigma(\theta)d\omega(t),\\
        Y_k &= C(\theta)X_k+v_k,\quad
        Z(t) = C_z(\theta)X(t),
\end{alignat}
\end{subequations}
where $X(t), U(t), D(t), Y_k$ and $Z(t)$ represent the deviation of the variables from the operating point, and the matrices $A(\theta)$, $B(\theta)$, $E(\theta)$, $C(\theta)$, and $C_z(\theta)$ are defined as
\begin{subequations}
\label{eq:ss_matrices}
\begin{alignat}{2}
        A(\theta) &=\frac{\partial}{\partial x}f(x_s,u_s,d_s, \theta), \quad 
        C_z(\theta) &&= \frac{\partial}{\partial x}h(x_s,\theta),\\ 
        B(\theta) &=\frac{\partial}{\partial u}f(x_s,u_s,d_s,\theta), \quad 
        C(\theta) &&=  \frac{\partial}{\partial x}g(x_s, \theta), \\
        E(\theta) &=\frac{\partial}{\partial d}f(x_s,u_s,d_s,\theta).
\end{alignat}
\end{subequations}

\subsection{State augmentation and offset-free estimation}
We augment the process models with integrating disturbance models such that the filters provide offset-free estimation \citep{Jorgensen:2007,Jorgensen:ACC:2007,Rawlings2003}.  
We model the disturbances as a stochastic process described by SDEs assuming the drift term to be constant, i.e., $dd(t) = 0dt+\sigma_d(\theta)d\omega_d(t)$,
with $ \sigma_d(\theta) = \mathrm{diag}([\sigma_{d,1},\: \sigma_{d,2},\: \sigma_{d,3},\: \sigma_{d,4}])$. The disturbance-augmented process model has almost the same structure as \eqref{eq:model_structure}; i.e. it has the redefined states, $x(t) := [x(t); d(t)]$, and the diffusion coefficients being $\sigma(\theta) := \mathrm{diag}([\sigma_1,\: \sigma_2,\: \sigma_3,\: \sigma_4,\: \sigma_{d,1},\: \sigma_{d,2},\: \sigma_{d,3},\: \sigma_{d,4}])$. The dynamical model augmented with the disturbances is $dx(t) = f(x(t),u(t),\theta) dt + \sigma(\theta) d\omega(t)$. This model and the corresponding measurement equation is used for the state estimation.

\section{State Estimation}
\label{sec:estimation}

We use a CD-EKF for parameter estimation in a ML estimation formulation and to estimate the states and disturbances in the NMPC \citep{Brok}. We use a CD-KF to estimate the states and disturbances in the LMPC. The state estimation is based on the model \eqref{eq:model_structure} augmented with a disturbance model.

\textit{Time-update}: The one-step prediction,
\begin{subequations}
\begin{alignat}{1}
        \hat{x}_{k|k-1} &= \hat{x}_{k-1}(t_{k}),\quad
        P_{k|k-1} = P_{k-1}(t_{k}),
\end{alignat}
\end{subequations}
is computed by numerical solution of
\begin{subequations}
\label{eq:timeupdate_CDEKF}
\begin{alignat}{3}
        \frac{d}{dt}\hat{x}_{k-1}(t) &= f( \hat{x}_{k-1}(t), u_{k-1}, \theta)  
        \label{eq:CDEKF_timeupdate}\\
        \frac{d}{dt}P_{k-1}(t) &= A_{k-1}(t)P_{k-1}(t)+P_{k-1}(t)A_{k-1}(t)'\nonumber\\
        &+\sigma(\theta) \sigma(\theta)',
\end{alignat}
\end{subequations}
for $t \in [t_{k-1}, t_k]$ with the initial conditions
\begin{equation}
    \hat{x}_{k-1}(t_{k-1}) = \hat{x}_{k-1|k-1},\quad
        P_{k-1}(t_{k-1}) = P_{k-1|k-1},
\end{equation}
where 
\begin{equation}
    A_{k-1}(t) = \frac{\partial}{\partial x}{f}(\hat{x}_{k-1}(t), u_{k-1}, \theta).
\end{equation}
 The ODEs are solved using the classical 4th order explicit Runge-Kutta method with 10 fixed time steps in each control interval.

\textit{Measurement-update:} The CD-EKF computes the current estimate, $\hat{x}_{k|k}$, and its covariance, $P_{k|k}$, based on the previous predicted estimate, $\hat{x}_{k|k-1}$, and covariance, $P_{k|k-1}$,
\begin{subequations}
\begin{alignat}{3}
        \hat{x}_{k|k} &= \hat{x}_{k|k-1}+K_ke_k,\\
        P_{k|k} &= (I-K_kC_k)P_{k|k-1}(I-K_kC_k)'+K_kRK_k',
\end{alignat}
\end{subequations}
where 
\begin{subequations}
\begin{alignat}{3}
    \hat{y}_{k|k-1} &= g(\hat{x}_{k|k-1}, \theta), \quad    && C_k = \frac{\partial}{\partial x}g(\hat{x}_{k|k-1}, \theta), \\
     e_{k} &= y_k-\hat{y}_{k|k-1},  \quad && R_{e,k} = R+C_kP_{k|k-1}C_k',\\
    K_k &= P_{k|k-1}C_k'R_{e,k}^{-1}. \quad && 
\end{alignat}
\end{subequations}

\begin{remark}[CD-KF]
The LMPC is based on a CD-KF. The CD-KF uses the innovation, $e_{k} = Y_k-C_k\hat{X}_{k|k-1}$ with $C_k = C(\theta)$ precomputed.
For the time-update $A_{k-1}(t) = A(\theta)$ and the linear model is used in (\ref{eq:CDEKF_timeupdate}).
\end{remark}

\section{A Maximum Likelihood Prediction-Error-Method}
\label{sec:MLPEM}
\begin{figure}[tb]
    \centering
    \centerline{\includegraphics[width=0.5\textwidth]{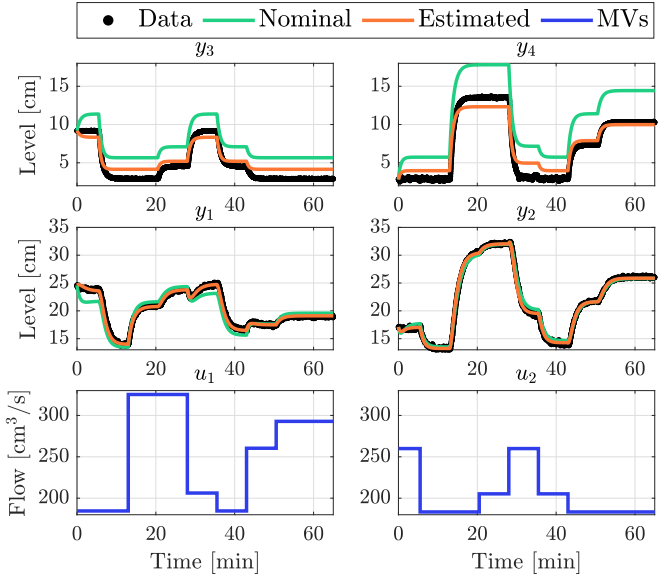}}
    \caption{Data used for estimation and simulations with nominal and estimated parameters.}
    \label{fig:estimation_data}
\end{figure}

\begin{figure}[tb]
    \centering
    \centerline{\includegraphics[width=0.5\textwidth, trim={0 0 0 0},clip]{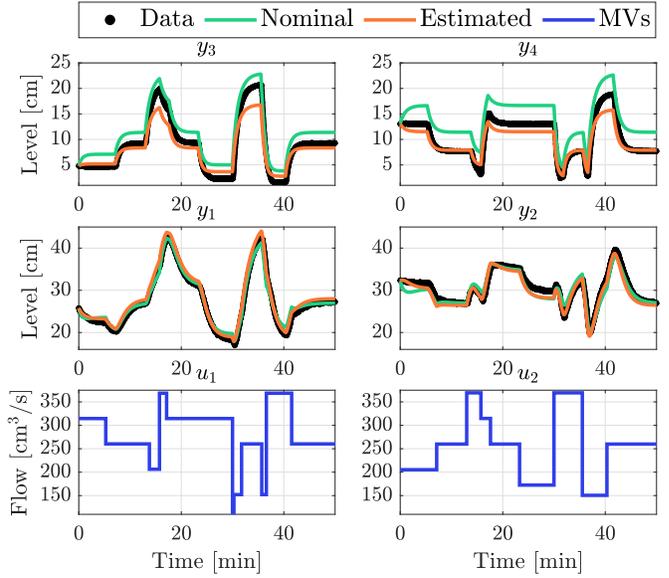}}
    \caption{Data used for validation and simulations with nominal and estimated parameters.}
    \label{fig:validation_data}
\end{figure}

Given a set of $N$ measurements and MVs,
\begin{subequations}
\begin{alignat}{3}
Y_N &= \begin{bmatrix}
    y_1,& y_2, &y_3,  & \dots, & y_N
    \end{bmatrix},\\
    U_N &= \begin{bmatrix}
    u_1,& u_2, &u_3,  & \dots, & u_N
    \end{bmatrix},
\end{alignat}
\end{subequations}
and given a model (\ref{eq:model_structure}), the maximum likelihood estimates of the parameter $\theta$ denoted $\theta_{ML}^*$, is the parameter vector that maximizes the likelihood function, $p(Y_N|\theta)$, i.e., the likelihood of obtaining the sequence of measurements in $Y_N$. We apply the rule for the product of conditional densities to the likelihood function,
\begin{align}
    p(Y_N|\theta) &= \prod_{k=1}^Np(y_k|\theta) = \prod_{k=1}^N p(e_k|\theta) 
    \\&= \prod_{k=1}^N \frac{1}{2\pi^{n_y/2}\sqrt{\det( R_{e,k})}}\exp\bigg(-\frac{1}{2}e_k' R_{e,k}^{-1}e_k\bigg),
    \label{eq:joint_prob_desnity_rule}
\end{align}
by assuming that the innovations are normally distributed, $e_k \sim N_{iid}(0,R_{e,k})$. $n_y = 4$ is the dimension of the measurement vector. The innovation, $e_k$, and its covariance, $R_{e,k}$, are computed using a CD-EKF.
We define the objective function $V_{ML}(\theta) =-\ln(p(Y_N|\theta))$,
\begin{equation}
     V_{ML}(\theta)= \frac{1}{2}\sum_{k=1}^{N}\bigg(\ln{\det( R_{e,k})}+e_k'R_{e,k}^{-1}e_k \bigg) + \frac{Nn_y}{2}\ln{2\pi},
\end{equation}
and the maximum likelihood estimate is computed as $\theta_{ML}^*~=~\arg\min ~ V_{ML}(\theta)$ \citep{Kristensen2004}.
We generate the estimation data by applying random step changes to the MVs. 
$\theta_{ML}^*$ is computed using this data and is presented in Table \ref{tab:all_parameters} together with the nominal parameters. The parameters $\rho$ and $g_a$ are not estimated. Figure \ref{fig:estimation_data} presents the input-output data used for estimation, i.e. the flows and the measured water levels. It also shows simulations using the nominal parameters and the parameters estimated from the data. We generate a second set of data for validation of the estimated parameters. Figure \ref{fig:validation_data} presents the validation data and simulations based on nominal parameters and the parameters estimated from the data in Figure \ref{fig:estimation_data}.

\begin{table}[tb]
\centering
\caption{Nominal and estimated parameters for \eqref{eq:model_structure}.}
\begin{tabular}{crrc}
\hline
Parameter ($\theta$) & Nominal & Estimated & Unit \\ \hline
$a_1$  & 1.131 & 1.006  & cm$^2$  \\
$a_2$  & 1.131 & 1.249  & cm$^2$\\
$a_3$  & 1.131 & 1.315& cm$^2$ \\
$a_4$  & 1.131 & 1.548 & cm$^2$ \\
$A_1$  & 380.133 & 379.837& cm$^2$ \\
$A_2$  & 380.133 & 378.034& cm$^2$\\
$A_3$  & 380.133 & 466.300& cm$^2$\\
$A_4$  & 380.133 & 523.122& cm$^2$\\
$\gamma_1$  & 0.350 & 0.260 & -- \\
$\gamma_2$  & 0.350 & 0.353 & --\\
$\sigma_{1}$ & - & 10.07 $\cdot 10^{-3}$ & g/$\sqrt{\mathrm{s}}$\\
$\sigma_{2}$   & - & 13.09 $\cdot 10^{-3}$ & g/$\sqrt{\mathrm{s}}$ \\
$\sigma_{3}$   & - & 12.50 $\cdot 10^{-3}$ & g/$\sqrt{\mathrm{s}}$\\
$\sigma_{4}$   & - & 16.62 $\cdot 10^{-3}$ & g/$\sqrt{\mathrm{s}}$\\
\hline
\end{tabular}

\label{tab:all_parameters}
\end{table}

We measure the goodness-of-fit (GOF) between data from the QTS and the water levels from the open-loop simulations of the system \eqref{eq:model_structure} by computing the averaged normalized root mean squared error,
\begin{equation}
    \mathrm{GOF} = \frac{1}{n_y} \sum_{i=1}^{n_y}\sum_{k=1}^{N} \bigg(1-\frac{||y_{i,k}-\tilde{y}_i(t_k)||}{||y_{i,k}-\mathrm{mean}(y_i)||}\bigg)100.
\end{equation}
$y_{i,k}$ and $\tilde{y}_i(t_k)$ for $i \in \{1, 2, 3, 4\}$ are data from the QTS and simulated water levels in tanks using \eqref{eq:meas_eq} without noise, respectively. Table \ref{tab:GOF_estimation_and_validation} presents the GOF for simulations using \eqref{eq:model_structure} with the nominal parameters and the estimates of the parameters. As expected, the GOF is significantly higher when using estimates of the parameters instead of the nominal parameters.

\begin{table}[b]
\centering
\caption{GOF for estimation and validation data.}
\begin{tabular}{llllll}
\hline
Parameters ($\theta$) &  Estimation GOF &  Validation GOF \\ \hline
 Nominal & 47.91\% & 57.28\%  \\
Estimated & 80.41\% & 74.20\%\\
\hline
\end{tabular}
\label{tab:GOF_estimation_and_validation}
\end{table}

\section{Control Algorithms}
\label{sec:control}
We present the three control algorithms in a descriptive manner. We denote the setpoints for the two bottom tanks as $\Bar{z}_k~=~\begin{bmatrix}
\Bar{z}_{1,k}; & \Bar{z}_{2,k}
\end{bmatrix}$ and the rate of change in the MVs as $\Delta u_k = u_{k+1}-u_{k}~=~\begin{bmatrix}
u_{1,k+1}-u_{1,k}; & u_{2,k+1}-u_{2,k}
\end{bmatrix}$. We implement the PID control system as two single-input single-output (SISO) loops. As pump 1 primarily influences tank 2 and 4 and pump 2 primarily influences tank 1 and 3, we choose the first loop to use $y_1$ and $\Bar{z}_1$ to compute $u_2$, and the other loop to use $y_2$ and $\Bar{z}_2$ to compute $u_1$. We include anti-windup in the SISO PID loops as the MVs will be constrained between upper and lower bounds. The PID loops are based on the description in \cite{Astrom1997}. The LMPC consists of an OCP based on the discrete-time linear state-space model of the system with input constraints and a CD-KF for estimating states and unmeasured disturbances \citep{Sazuan_MPC}. The NMPC consists of an OCP with input constraints and a CD-EKF for estimating states and unmeasured disturbances. The OCP contains the continuous-time deterministic nonlinear model of the system, and we discretize it using a direct multiple shooting formulation implemented with CasADi \citep{Casadi}. The objective functions in the LMPC and NMPC penalize the quadratic tracking errors between setpoints and the CVs using the weight matrix $Q$, and the quadratic rate of change in the MVs using the weight matrix $S$. The control algorithms are all implemented using the Python programming language and a sampling time, $T_s$, of 5~s is chosen for all three control algorithms.

\subsection{Tuning of Controllers}

To tune the SISO PID loops, $(z_1,u_2)$ and $(z_2,u_1)$, we compute the transfer functions from MVs to CVs of the QTS from \eqref{eq:linear_state_space_model} as
\begin{equation}
    G(s) = \begin{bmatrix}
g_{11}(s) & g_{12}(s) \\ 
g_{21}(s) & g_{22}(s)
\end{bmatrix} = C_z(\theta) (s I - A(\theta))^{-1} B(\theta).
\end{equation}
$g_{12}(s)$ and $g_{21}(s)$ are second order systems in the form
\begin{equation}
\label{eq:transfer_functions}
    g(s) = \frac{k}{(\tau_1s+1)(\tau_2s+1)},
\end{equation}
where $\tau_1 \geq \tau_2$ are time constants, and $k$ is the steady-state gain.
We compute the proportional gain, $K_p$, the integrator time constant, $\tau_i$, and the derivative time constant, $\tau_d$, for each SISO PID loop, using the simple internal model control (IMC) rules,
\begin{subequations}
\begin{alignat}{3}
 \Tilde{K}_p &= \frac{\tau_1}{k T_c}, \,\Tilde{\tau}_i = \min(\tau_1, 4T_c),\, \Tilde{\tau}_d = \tau_2,\\
 K_p &= \Tilde{K}_p\alpha, \,\tau_i =  \Tilde{\tau}_i\alpha,\, \tau_d = \frac{\Tilde{\tau}_d }{\alpha}, 
\end{alignat}
\end{subequations}
with $\alpha = 1+\frac{\Tilde{\tau}_d}{\Tilde{\tau}_i}$ \citep{inbook_Skogestad}. We choose the tuning parameter $T_c = 50$ for both PID loops in the PID control system. For the LMPC and NMPC, we choose the weight matrices and number of control and prediction steps, $N_c$, as
\begin{equation}
    Q = \mathrm{diag}([10,\:10]),\quad
        S = \mathrm{diag}([1,\:1]),\quad
        N_{c} = 160.
    \label{eq:MPC_tuning}
\end{equation}
For a sampling time of 5 s the prediction horizon for the LMPC and NMPC is $N_{c}T_s = 13.33$ min which is a sufficiently long horizon when considering time response of the QTS. The disturbance-augmented CD-KF and CD-EKF are tuned identically with the diffusion coefficients and measurement noise covariance in Table \ref{tab:KF_tuning}.

\begin{table}[tb!]
\centering
\caption{Estimated diffusion coefficients, $\sigma(\theta)$, and measurement noise covariance, $R$.}
\label{tab:KF_tuning}
\begin{tabular}{llll}
\hline
$\sigma_1$ & $\sigma_2$ & $\sigma_3$ & $\sigma_4$  \\ 
 7.25 & 14.92 & 8.98 & 14.50  \\ \hline
$\sigma_{d,1}$ & $\sigma_{d,2}$ & $\sigma_{d,3}$ & $\sigma_{d,4}$  \\ 
0.47 & 3.08 & 3.92 & 3.42 \\ \hline
 $r_{1}^2$ & $r_{2}^2$ & $r_{3}^2$ & $r_{4}^2$  \\ 
1.44$\cdot 10^{-2}$ & 1.34$\cdot 10^{-2}$ & 1.00$\cdot 10^{-5}$ & 1.00$\cdot 10^{-5}$  \\ \hline
\end{tabular}%

\end{table}

\section{Experimental Setup and Results}
\label{sec:experiments}

\begin{figure}[tb]
    \centering
    \centerline{\includegraphics[width=0.5\textwidth]{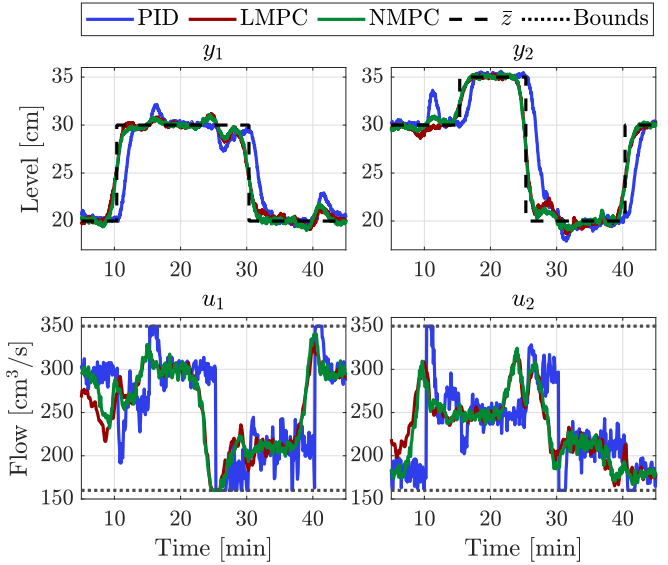}}
    \caption{Data for experiments of PID, LMPC, and NMPC.}
    \label{fig:comparison_ML}
\end{figure}

\begin{figure}[tb]
    \centering
    \centerline{\includegraphics[width=0.5\textwidth ]{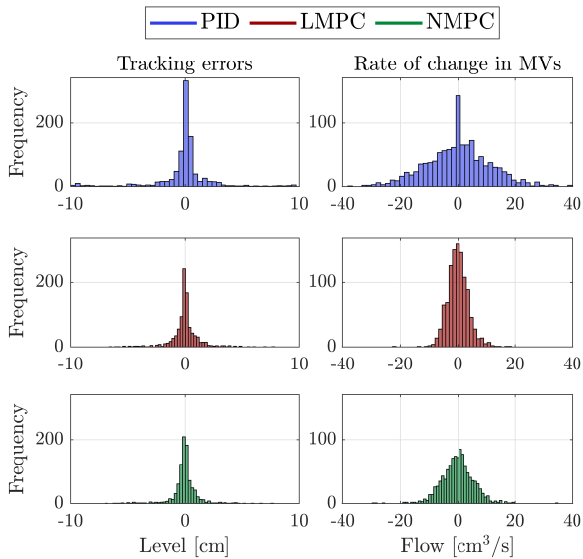}}
    \caption{Histograms of tracking errors and rate of change in MVs for the data in Figure \ref{fig:comparison_ML}.}
    \label{fig:histograms}
\end{figure}

We conducted a closed-loop experiment for each of the three control strategies for a physical setup of the QTS. Predefined time-varying setpoints with no steps occurring simultaneously for both tanks were tested. The MVs were bounded between $160\: \mathrm{cm}^3/\mathrm{s} \leq u_{i}(t) \leq 350\: \mathrm{cm}^3/\mathrm{s}$ for $i \in \{1, 2\}$. The chosen operating point for the linear models used for the PID and the LMPC design, was $ u_s~=~\begin{bmatrix}
300\:\mathrm{cm}^3/\mathrm{s}; & 300\:\mathrm{cm}^3/\mathrm{s}
\end{bmatrix}$, $d_s~=~\begin{bmatrix}
0\:\mathrm{cm}^3/\mathrm{s}; & 0\:\mathrm{cm}^3/\mathrm{s};  & 0\:\mathrm{cm}^3/\mathrm{s};  & 0\:\mathrm{cm}^3/\mathrm{s}
\end{bmatrix}$, 
and $x_s$ was computed solving $0 = f(x_s, u_s, d_s, \theta)$.
The implemented control algorithms received the measured water levels from the QTS, $y_k$, and applied the MVs to the QTS, $u_k$, through an open platform communications unified architecture (OPC-UA) connection with the physical setup, and data from the experiments were stored using an SQL database system.
Figure \ref{fig:comparison_ML} presents the data from the experiments. Figure \ref{fig:histograms} presents histograms of the tracking errors and rate of change in MVs.
We define the measured tracking errors using the measurements of the water levels in the bottom tanks as
\begin{equation}
    \bar{e}_k = \bar{z}_k - \begin{bmatrix} y_{1,k}; & y_{2,k}\end{bmatrix}.
\end{equation}
We apply different norms to the tracking error and the rate of change in the MVs to measure the performance of the controllers, i.e., we compute the normalized integral squared error (NISE), the normalized integral absolute error (NIAE), and the normalized integral squared rate of change in MVs (NIS$\Delta$U) as
\begin{subequations}
\begin{alignat}{1}
         \mathrm{NISE} &= \frac{1}{N}\sum_{k = 1}^N\left \|   \bar{e}_k\right \|^2_2, \quad
         \mathrm{NIAE} = \frac{1}{N}\sum_{k = 1}^N\left \|   \bar{e}_k\right \|_1,\\
       \mathrm{NIS\Delta U} &= \frac{1}{M-1}\sum_{j = 1}^{M-1}\left \|   \Delta u_j\right \|^2_2,
\end{alignat}
\end{subequations}
where $N$ and $M$ are the number of data points of the water level and of the MVs computed by the controller, respectively. Table \ref{tab:ML_performance_control} presents the computed performance metrics.

\begin{table}[tb!]
\centering
\caption{Computed performance measures of PID, LMPC, and NMPC for the data in Figure \ref{fig:comparison_ML}.}
\begin{tabular}{lrrr}
\hline
Control & NISE & NIAE & NIS$\Delta$U  \\ \hline
PID  & 9.063 & 1.459 & 249.079 \\
LMPC & 1.637 & 0.728 & 12.089 \\
NMPC & 1.423 & 0.647 & 28.674\\
\hline
\end{tabular}
\label{tab:ML_performance_control}
\end{table}

Compared with the PID, the LMPC and the NMPC significantly improve the performance of the QTS when considering the tracking of predefined setpoints. The rate of change in the MVs is also reduced considerably when using the LMPC and NMPC instead of the PID. This is documented in Figure \ref{fig:histograms} that also shows that the tracking error outliers are removed for the LMPC and NMPC. The NMPC only provides slightly improved tracking errors compared with the LMPC, but the rate of change in the MVs is larger for the NMPC. Consequently, the MPCs do improve the performance compared with a PID controller. For this case study, the NMPC and LMPC provide similar performance.

\section{Conclusions}
\label{sec:conclusion}

In this paper, we present a comparative study of the performance for implementations of a PID controller with model-based tuning, an LMPC, and an NMPC applied to a physical setup of a quadruple tank system. The model for these controllers is estimated using a prediction-error-method. We apply a predefined time-varying setpoint trajectory to compare the controllers. Based on the tracking error and input variation, LMPC and NMPC provide better performance than the PID controller. LMPC and NMPC have similar performance.


%
%

\bibliographystyle{chicago}
\section{References}


\def\refname{}
\def\bibsection{}

\bibliography{ref/References.bib}%


%
%
%
%
%


\end{document}